\documentclass[review]{elsarticle}

\usepackage{lineno,hyperref}
\usepackage{breqn}
\usepackage{amsfonts}
\usepackage{siunitx}
\usepackage{caption}
\usepackage{subcaption}
\usepackage{float}
\modulolinenumbers[5]
\journal{NIM-A}

\begin{document}

\begin{frontmatter}

\title{A neural network for beam background decomposition in Belle~II at SuperKEKB}

\author[instGoe]{B.~Schwenker\corref{mycorrespondingauthor}}
\ead{benjamin.schwenker@phys.uni-goettingen.de}
\author[instGoe]{L.~Herzberg}
\author[instGoe]{Y.~Buch}
\author[instGoe]{A.~Frey}
\author[instHawaii]{A.~Natochii}
\author[instHawaii]{S.~Vahsen}
\author[instKEK,instSOKENDAI]{H.~Nakayama}

\address[instGoe]{II. Physikalisches Institut, Georg-August-Universit\"at G\"ottingen, 37073 G\"ottingen, Germany}

\address[instHawaii]{University of Hawaii, Honolulu, Hawaii 96822, USA}

\address[instKEK]{High Energy Accelerator Research Organization (KEK), Tsukuba 305-0801}

\address[instSOKENDAI]{SOKENDAI (The Graduate University for Advanced Studies), Hayama 240-0193}

\cortext[mycorrespondingauthor]{Corresponding author}

\begin{abstract}
We describe a neural network for predicting the background hit rate in the Belle~II detector produced by the SuperKEKB electron-positron collider. The neural network, BGNet, learns to predict the individual contributions of different physical background sources, such as beam-gas scattering or continuous top-up injections into the collider,  to Belle~II sub-detector rates. The samples for learning are archived $\SI{1}{\hertz}$ time series of diagnostic variables from the SuperKEKB collider subsystems and measured hit rates of Belle~II used as regression targets. We test the learned model by predicting detector hit rates on archived data from different run periods not used during training. We show that a feature attribution method can help interpret the source of changes in the background level over time.
\end{abstract}

\begin{keyword}
Belle~II\sep SuperKEKB \sep  Beam background \sep Neural networks \sep Nonlinear regression \sep Machine learning for accelerators 
\end{keyword}

\end{frontmatter}


\section{Introduction}


The Belle~II experiment at SuperKEKB, an asymmetric electron-positron collider, aims to collect an unprecedented data set of $50~{\rm ab}^{-1}$ for high precision studies of the flavour sector and to search for physics beyond the Standard Model. SuperKEKB, located at KEK (Tsukuba, Japan), collides $\SI{7}{GeV}$ electrons with $\SI{4}{GeV}$ positrons at a center of mass energy of $\SI{10.58}{GeV}$ which corresponds to the rest mass of the $\Upsilon(4S)$ resonance. SuperKEKB has reached a world-record luminosity of $\SI{4.7e34}{cm^{-2}.s^{-1}}$ for a vertical betatron function of $\beta^{*}_{\rm y}=\SI{1.0}{mm}$ at the interaction point (IP) in summer 2022. In order to collect the planned data set in the next ten years, the target is to reach a peak luminosity of $\SI{6.3e35}{cm^{-2}.s^{-1}}$ by further increasing the beam currents and reducing the beam-size at the interaction point by squeezing the betatron function down to 
$\beta^{*}_{\rm y}=\SI{0.3}{mm}$. The Belle~II experiment is a general-purpose particle detector built around the interaction point of SuperKEKB where the electrons and positrons collide. It is composed of several subsystems for tracking, particle identification and calorimetry. A detailed review of the SuperKEKB collider and its subsystems can be found in Refs.~\cite{Ohnishi2013,Ishibashi2020,SUETSUGU2008153,Mitsuka2020}. Reviews for the design of the Belle~II detector can be found in Refs.~\cite{Abe:2010sj, ADACHI201846}.


Beam backgrounds seen by the Belle~II detector originate from beam particles lost near the collision point. Beam particles which deviated from the nominal orbit are eventually lost by hitting the beam pipe inner wall or other machine apparatus. If their loss position is close to the interaction point, generated shower particles might reach the Belle~II detector. The ionizing and non-ionizing irradiation from beam backgrounds sets limits on the lifetime of subsystems installed in the  Belle~II detector. Particles in background showers generate fake hits which overlay hits from triggered signal collisions posing a challenge to the event reconstruction software~\cite{basf2}. An efficient operation of the Belle~II experiment requires to reach the target luminosity but also to keep beam backgrounds below detector limits. 

In order to control and mitigate particle losses near the interaction point, a set of moveable beam collimators is installed around the high energy electron ring (HER) and the low energy positron ring (LER). There are two main types of collimators: KEKB type collimators with one jaw and SuperKEKB type collimators with two jaws. SuperKEKB type collimators can be set asymmetric, i.e. the width of the inner jaw and the width of the outer jaw can be set differently. More details on the moveable collimators installed in SuperKEKB can be found in Ref.~\cite{Ishibashi2020}. The collimators play a central role in the mitigation of backgrounds through stopping stray particles in the beam halo. Therefore, the monitored position of the collimator jaws relative to the beam center forms an important group of variables to predict backgrounds.   


The contribution of this work is a neural network, BGNet, for the prediction of the hit rate of individual background sources in the Belle~II  detector. The proposed network is structured into submodels for individual background sources each exploiting heuristics from scattering theory to facilitate a physically sensible decomposition of the total hit rate. The neural network is trained on archived time series of the hit rate of a selected Belle~II sub-detector as the regression target and multiple input time series of selected variables monitoring different subsystems of the SuperKEKB collider. After training, the neural network can predict all background components at any time given the values of the selected collider input variables. We propose BGNet as a diagnostic tool to provide real time predictions of background components seen in Belle~II detectors during the operation of SuperKEKB.

In recent years, feature attribution methods~\cite{Sundararajan, Erion}, like expected gradients, have emerged as a tool to quantify how much individual input variables contribute to the prediction of a model relative to a baseline. We demonstrate how expected gradients can be used to find the origin of a change in the background level over time.

\section{Background types}

A number of different physical processes contribute to the rate of particle losses near the collision point. A detailed reference of the physical processes can be found in Ref.~\cite{Wolski2014}. For the case of Belle~II, the most important background sources found during previous studies, see Ref.~\cite{Andrii2022}, are:

\paragraph{Beam-gas background}

Beam-gas background originates from the interaction between beam particles and the residual gas atoms/molecules in the evacuated beam pipe. Beam-gas Coulomb scattering changes the direction of the beam particles, and beam-gas Bremsstrahlung scattering reduces the energy of the beam particles. The beam-gas scattering rate is proportional to the vacuum pressure in the beam pipe and the beam current. Simulation studies show that the loss rate in Belle~II can be reduced by vertical collimators with a narrow aperture.

\paragraph{Touschek background}

Touschek scattering occurs when two particles in the same bunch approach each other closely enough that they are deflected by an angle leading to a significant transfer of momentum from a transverse to the longitudinal direction. This Coulomb scattering between two particles increases the longitudinal momentum of one particle while it decreases the longitudinal momentum of the other.  After the scattering event, the energy deviation of one or both particles may be outside the energy acceptance of the collider. The Touschek scattering rate is proportional to the beam current squared and inversely proportional to the number of bunches and the beam size. It is expected that the Touschek background is sensitive to the aperture of horizontal collimators installed around the ring.    

\paragraph{Injection background}
    
Beam losses due to Touschek and beam-gas scattering limit the beam lifetime of SuperKEKB well below one hour. To allow stable operations for long periods, it is necessary to perform top-up injections following a betatron injection scheme~\cite{Ohnishi2013} during physics data taking. The newly injected bunch is perturbed and oscillates in the horizontal plane around the main stored beam. It causes high background rates in the Belle~II detector for a few milliseconds after injection. During a physics run, top-up injections with up to $\SI{25}{Hz}$ per beam are used once the beam current falls below a limit and will be paused for multiple seconds once a target current is reached. The temporary pausing of injections is indicated by a binary variable called beam gate status.  
    
\paragraph{Luminosity background}

Luminosity background originates from electron-positron collisions at the interaction point inside the Belle~II detector. The dominant processes are radiative Bhabha scattering $e^+e^- \to e^-e^-\gamma$ and the two photon process $e^+e^- \to e^+e^-e^+e^-$. The hit rate from luminosity background is proportional to instantaneous luminosity and becomes dominant at the target luminosity of SuperKEKB, which is about 30~times higher than the record of KEKB~\cite{Abe2013}.

Motivated by the theory of beam dynamics in electron positron storage rings \cite{Wolski2014}, the hit rate $\mathcal{O}$ of Belle~II subsystems due to beam-gas background can be approximated by the formulae 

\begin{equation}
    \mathcal{O}_{\rm Beam\text{-}gas,H} = S_1 \times I_{\rm H} P_{\rm H} = S_1 \times G_1,
    \label{eq1}
\end{equation}

\begin{equation}
    \mathcal{O}_{\rm Beam\text{-}gas,L} = S_2 \times I_{\rm L}P_{\rm L} = S_2 \times G_2,
    \label{eq2}
\end{equation}
where $I$ is the stored beam current and $P$ is the effective residual gas pressure seen by the beam in the center of the evacuated beam pipe. The subscripts H or L are used for variables related to the high energy electron ring (HER) or the low energy positron ring (LER) respectively. The coefficient $S_1$ ($S_2$) parametrizes the sensitivity to the beam-gas background from the HER (LER). For the determination of the effective residual pressure $P$ from the readings of the pressure gauges placed around the ring, we follow the approach described in detail in Ref.~\cite{Andrii2022}. 

The hit rate due to Touschek scattering can be approximated by the formulae

\begin{equation}
    \mathcal{O}_{\rm Touschek,H} = S_3 \times \frac{I_{\rm H}^{2}}{n_{\rm b,H}\sigma_{\rm x,H}\sigma_{\rm y,H}\sigma_{\rm z,H}} = S_3 \times G_3,
    \label{eq3}
\end{equation}

\begin{equation}
    \mathcal{O}_{\rm Touschek,L} = S_4 \times \frac{I_{\rm L}^{2}}{n_{\rm b,L}\sigma_{\rm x,L}\sigma_{\rm y,L}\sigma_{\rm z,L}} = S_4 \times G_4,
    \label{eq4}
\end{equation}
and depends on the bunch volume $\sigma_{\rm x}\sigma_{\rm y}\sigma_{\rm z}$ and the number of bunches $n_{\rm b}$ stored in the collider. The coefficient $S_3$ ($S_4$) is the sensitivity to the Touschek background from the HER (LER).      

We model the contribution of top-up injections to the hit rate by 

\begin{equation}
    \mathcal{O}_{\rm Inj, H} = S_5 \times G_5,
    \label{eq5}
\end{equation}

\begin{equation}
    \mathcal{O}_{\rm Inj, L} = S_6 \times G_6,
    \label{eq6}
\end{equation}
where $G_5$ ($G_6$) is an engineered injection heuristic and $S_5$ ($S_6$) describes the sensitivity of the hit rate to top-up injections into the HER (LER). We take the injection heuristic $G_5$ ($G_6$) to be unity when the product of the average injected charge into the HER (LER) and the HER (LER) beam gate status is positive and zero elsewhere. The injection sensitivity $S_5$ is expected to scale with the product of the repetition rate of injections $f_{\mathrm{Rep}}$, the average injected charge $Q_{\mathrm{Inj}}$  and the inefficiency of injections measured as the fraction of lost charge of the injected bunch in the first 100 turns in the collider. The fraction of charge losses near Belle~II will likely depend on other variables like the aperture of movable collimators. We use this expectations later by making these variables input features to the neural network model for background prediction.

The luminosity background scales linearly with the measured luminosity $\mathcal{L}$

\begin{equation}
    \mathcal{O}_{\rm Lumi} = S_7 \times \mathcal{L} = S_7 \times G_7,
    \label{eq7}
\end{equation}
where $G_7$ is the measured luminosity and $S_7$ is the sensitivity to the luminosity background. The formula for the total predicted hit rate of the Belle~II detector is
\begin{equation}
    \mathcal{O} = \sum_{i=1}^{7}   S_i \times G_i  + S_8,
    \label{eq8}
\end{equation}
where $S_8$ is a detector specific pedestal measurable when no beam is stored in the collider. For many Belle~II sub-detectors, the pedestal $S_8$ is stable over time.

\section{Background prediction}

In order to use Eq.~\eqref{eq8} for hit rate prediction due to backgrounds, we need to know the value of all the variables on the right hand side. The luminosity $\mathcal{L}$ and all variables needed to compute the coefficients $G_i$ are sampled at a frequency of $\SI{1}{\hertz}$ from the EPICS  Archiver Appliance \cite{EPICS_Archiver} of the Belle~II slow control system \cite{EPICS}. Finding the correct value of the sensitivity $S_i$ for each of the eight background sources is a more difficult task.

The traditional approach of the Belle~II collaboration was to measure the sensitivities to beam-gas and Touschek backgrounds during dedicated background study campaigns conducted once or twice a year \cite{Andrii2022}. The idea was to store a beam  only in one ring at a time and to record the decay of the beam current after pausing injections. By recording data from multiple single-beam decays, each decay differing only in the number of stored bunches, it is possible to disentangle beam-gas and Touschek backgrounds. For the same initial stored current, the density of particles per bunch and therefore the rate of Touschek scattering varies only with the number of bunches. Estimated values for the storage background sensitivities $S_1$,...,$S_4$ can be obtained from a least squares fit of the measured hit rates during single  beam decays against a simplified background model 

\begin{equation}
    \mathcal{O} = \sum_{i=1}^{4}   S_i \times G_i + S_{8},
    \label{eq9}
\end{equation}
because top-up injections are paused during beam decays and the luminosity is zero. The sensitivity to luminosity was obtained in a second step using data with colliding beams at a luminosity of   $\SI{2.6e34}{cm^{-2}.s^{-1}}$. The idea is to fit the difference of the measured hit rate and the sum of beam-gas and Touschek backgrounds with a linear model $S_7 \times \mathcal{L}$. Bias from injection backgrounds was avoided by ignoring data samples during or near periods with top-up injections into either the HER or LER. Measured sensitivities were published for many Belle~II sub-detectors~\cite{Andrii2022,LEWIS201969,Liptak2021}.

The extracted sensitivities from background study campaigns are a valuable information for the machine learning approach to background prediction. Training such a model with data from previous weeks of collider operation may contain insufficient information to constrain all background sensitivities. In this case it seems best to formulate the model in a way that neural network based models for background sensitivities can be initialized with values found during the latest background study day.

\section{Neural network for background prediction}

The measured detector hit rate seen at Belle~II during SuperKEKB operation may deviate from the expectations derived from background study data. For example, the sensitivity to storage backgrounds may have changed because the aperture of moveable collimators was adjusted. Or the amplitude of the injection background may have worsened due to an increase in the injection repetition frequency or the injected charge per bunch or other reasons. A large number of collider operation related diagnostics are available on the SuperKEKB archiver, timestamped with a frequency of $\SI{1}{\hertz}$. The space of diagnostic variables at a time forms a high dimensional feature space. Deep learning~\cite{bengio2016} with neural networks offers a way to learn maps from the space of collider features to the background sensitivities of Belle~II from archived data.  

\begin{figure}[!ht]
\centering
\includegraphics[width=0.8\linewidth]{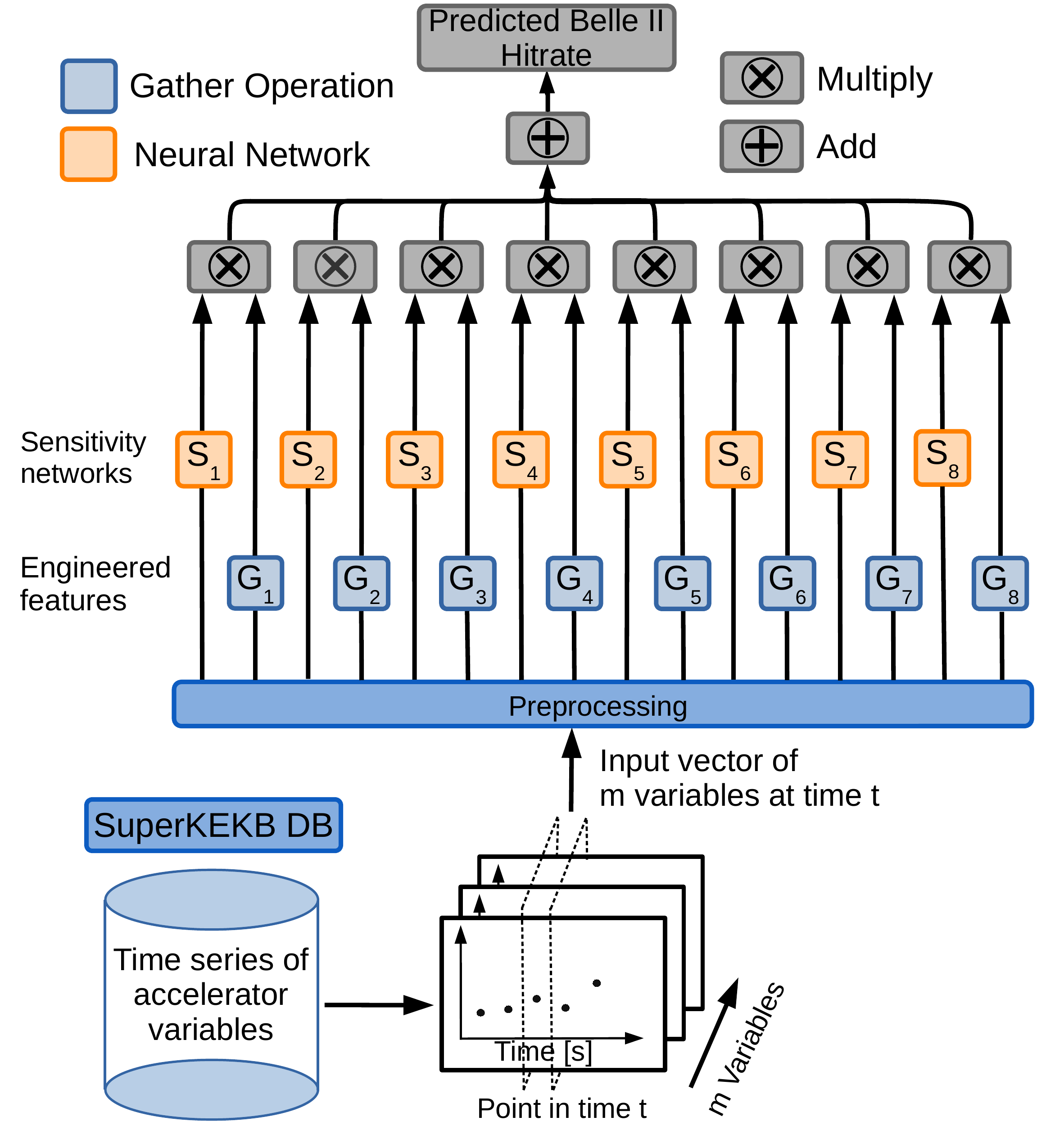}
\caption{\label{fig1}Flow diagram for BGNet: The output of the network is the total predicted background at a time $t$ for a Belle~II sub-detector.} 
\end{figure}

Figure~\ref{fig1} shows a flow diagram of BGNet with its different submodels and their connections. At the bottom, we have the SuperKEKB database (DB) with a collection of timestamped accelerator related diagnostics. The preprocessing provides a $m+8$ dimensional input vector $x(t)$ with accelerator variables as input to the neural network. The preprocessing includes the selection of variables from the DB for the queried time $t$, a time delay correction, the computation of engineered features and a scaling of all input features. The input array $x(t)$ is fed into a neural network to compute the predicted hit rate of a Belle~II sub-detector as output. The blocks $S_1$ to $S_8$ in Fig.~\ref{fig1} are fully connected feed forward networks and have a one dimensional positive output, i.e. the sensitivity for background type $i$ given the input vector $x$. The last eight entries in the input array are the engineered features defined in Eq.~\eqref{eq1} to Eq.~\eqref{eq7}\footnote{The feature $G_8$ for the detector pedestal is always one.}.
The eight sensitivities $S_i(x)$ are multiplied with the eight features $G_{i}(x)$ to yield the eight components of the background decomposition. The final step is a sum over the decomposition yielding the total predicted background as output. The background sensitivities and the decomposition are available from the intermediate network layers. 

\begin{figure}[!ht]
\centering
\includegraphics[width=0.8\linewidth]{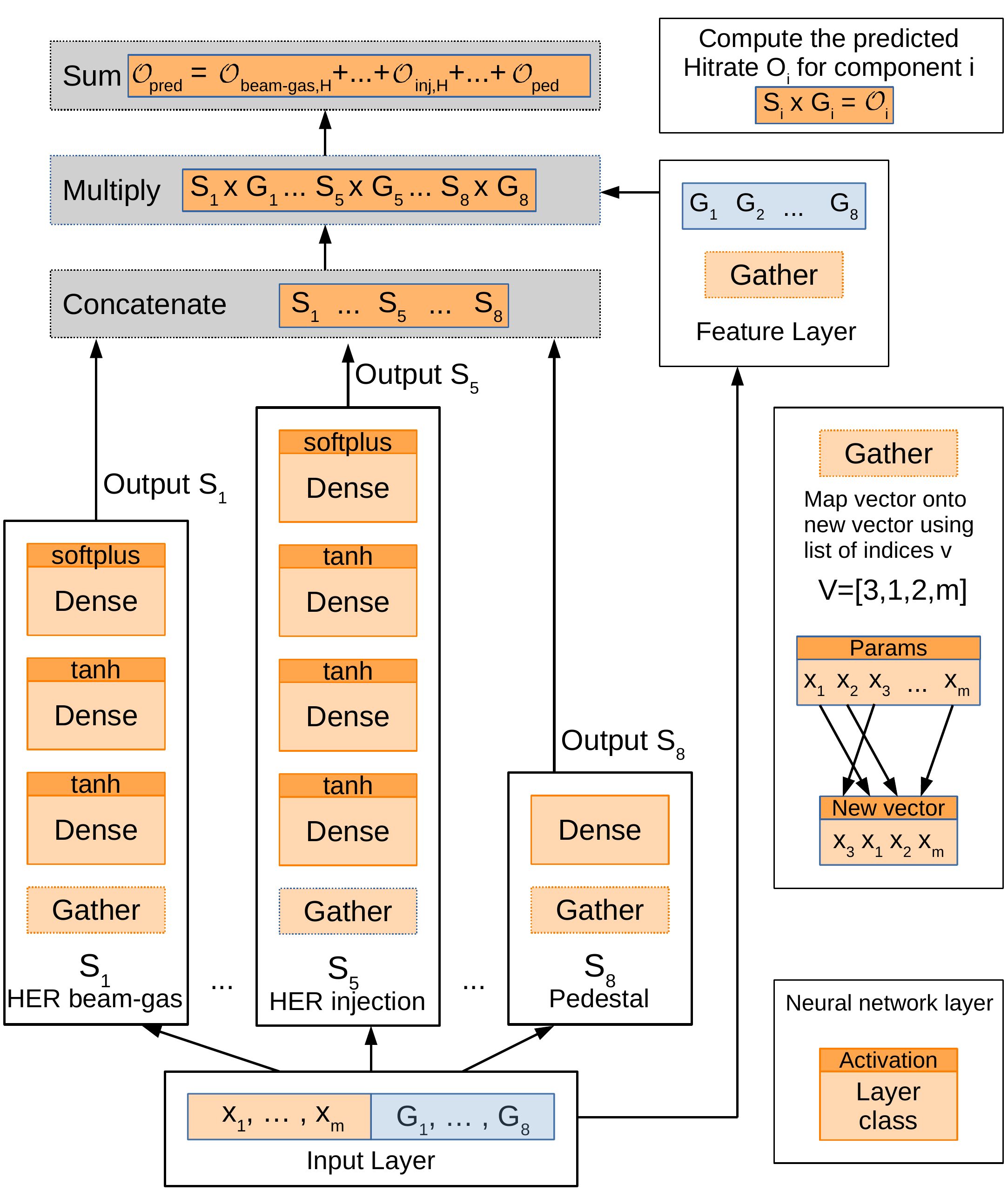}
\caption{\label{fig2} Flow diagram for BGNet with more details on the structure of sub networks. A dense layer is defined as a layer of neurons in which each neuron receives input from each neuron of the previous layer. The gather operation maps a vector onto a new vector using a list of indices. The activation function ``tanh'' refers to $f(x) = \tanh{x}$, while the activation function ``softplus'' refers to $f(x) = \ln(1+\exp(x))$.}
\end{figure}

Figure \ref{fig2} provides more details on the structure of the neural network. We know a priori that certain input features may be predictive for some sensitivities but cannot be causally related to others. For example  the apertures of collimators in the HER ring cannot have a direct causal relation to the background sensitivity for the LER beam-gas background. In a first order approximation, it may be sufficient to predict the LER beam-gas sensitivity from the aperture of vertical LER collimators and the LER Touschek sensitivity from the aperture of horizontal collimators.  In order to exploit such a priori information, we decided to create a list $V_i$ of selected input features for each sensitivity network. The common structure of the sensitivity network consists in projecting the full input array $x$ to the $m_i$ dimensional subspace of selected features $V_i$ and to feed this projected vector $x_i$ into a fully connected feed forward network. The network $\mathcal{S}_7$ and $\mathcal{S}_8$ form a special case since the sensitivity to the luminosity background and the detector noise should be constant and independent of accelerator conditions.  We realize each of these networks by a single linear neuron connected to constant inputs $x_7=x_8=1$.

The training objective for BGNet is to minimize the mean absolute error between the measured hit rate and the total predicted hit rate, i.e. the sum of all predicted background components and the detector pedestal.  We use the TensorFlow/Keras machine learning framework~\cite{tensorflow2015-whitepaper,chollet2015keras} to implement BGNet. The full source code for BGNet is open sourced and available online~\cite{BGNet_software}. Table~\ref{tab:network_parameters} summarizes hyperparameters of the neural network layers and the training procedure.

\begin{table}[h!]
    \centering
    \begin{tabular}{c|c}
        \hline\hline
        Number of hidden layers. & $S_{\text{1 to 4}}$: 2 \hspace{5px} $S_{\text{5 and 6}}$: 3\\
        Number of hidden units & $S_{\text{1 to 4}}$: 8 \hspace{5px} $S_{\text{5 and 6}}$: 32\\
        Activation function &  Hidden Layers: $f(x)=tanh(x)$\\ 
            &Output Layer: $f(x)=ln(1+exp(x))$\\
        Number of inputs & $S_1$: 9\hspace{5px}$S_2$: 15\hspace{5px} $S_3$: 8\hspace{5px}$S_4$: 12\hspace{5px} $S_5$: 33\hspace{5px} $S_6$: 27\\
        Loss & Mean absolute error\\
        Batch size & 32\\
        Optimizer & Adam~\cite{Kingma2014} \\
        Weights and bias initialization & Glorot uniform~\cite{pmlr-v9-glorot10a}\\
        \hline\hline
    \end{tabular}
    \caption{Table of hyper parameters for BGNet.}
    \label{tab:network_parameters}
\end{table}

\subsection{Selection of training data. }

The data source for training BGNet is the EPICS Archiver Appliance~\cite{EPICS_Archiver}. The training setup for BGNet requires the selection of a time window, ranging from a few days to multiple weeks, and the specification of the regression target, the hit rate of a Belle~II sub-detector, and a set of selected input features for sub models. After downloading the data from the EPICS Archiver Appliance, we apply a mask to select valid data for training. The mask selects data where the sub-detector was powered and operational, and a beam current $>10$mA is stored in at least one ring. Valid data was split into training and validation sets before the training. The training data is typically chosen to be the archived data from the latest few weeks of data taking with Belle~II. The key idea is that the model learns by following archived time series during normal day to day operation (i.e. physics run) and machine tuning to observe the effect of changed inputs on the measured hit rate.  

\subsection{Selection of input variables}

The selection of a list of input variables $V_i$ for predicting the different background sensitivities is a crucial step. Currently, the choice is guided by expert opinion seeking a plausible causal mechanism that a SuperKEKB variable can change the sensitivity of a certain beam background. For example, the aperture of vertical LER collimators is expected to influence the LER beam-gas background sensitivity but certainly not the storage backgrounds from the HER. In this sense, the selection of input variables provides a means to bias the model to learn causally relevant patterns. Candidate variables proposed by experts were tested by Bayesian hyperparameter optimization and by ranking them with feature attribution methods, see section~\ref{sec:explaining_bgnet}. The number of selected input variables is shown in Tab.~\ref{tab:network_parameters}.

\subsection{Preprocessing of data}

All input variables and the measured hit rate are scaled by subtracting the median and scaling by the percentile range between the $90$th and $10$th percentile. The centering of values was omitted for the measured hit rate and the engineered features $G_{1}$ to $G_{8}$. Using the scaled measured hit rate for computing the training loss improves the convergence of training since the raw hit rates from the archiver can be very big numbers. The center and scale parameters are computed on valid training data and are deployed along with the trained model. 

The last step of the preprocessing is the correction of an empirically observed time delay between the measured hit rate $\mathcal{O}(t)$ used as regression target and the accelerator input features $x(t)$. We parametrize the delay correction by a single delay shift $d=i + w$, given in seconds, where $i$ is the integer part and $w$ is the non integer part of $d$. The array, used as input by the network to predict the hit rate at time $t$, is the weighted mean of the inputs at i and (i+1) seconds in the past. 

\begin{equation}
    x^{\prime}(t) = (1-w)\cdot x(t-i) + w\cdot x(t-(i+1))
    \label{eq10}
\end{equation}
The variable $w$ is used as a weight to interpolate between two neighboring samples. The delay shift is treated as a hyperparameter of the model and the search of its optimal value is discussed in the results section.

\subsection{Model training}

During the operation of the collider we have moments where top-up injection is paused in both rings. During those moments, the measured hit rate is caused only by storage backgrounds, luminosity background and the detector pedestal. We find such decay data by looking only at moments where the product of the beam gate status and the injected charge is zero for HER and LER, i.e. $G_5=0$ and $G_6=0$. As can be seen from Fig.~\ref{fig1}, this implies that injection backgrounds cannot contribute to the output of BGNet. The fraction of decay data varies between a few percent on some days to up to 20\% on others. In order to disentangle the contributions of injection backgrounds on the one hand and storage plus luminosity on the other hand, it is important that the model describes decay data and non-decay data equally well. We apply random oversampling~\cite{imbalance_learn} to balance the ratio between decay and non-decay samples in the training data. We randomly select and duplicate examples of decay data to achieve a balanced 50:50 ratio between decay data and non decay data during the training. The oversampling of decay data is applied day by day to make sure the 50:50 balance is achieved for any day.

Disentangling the contribution of beam-gas, Touschek and luminosity backgrounds to the total background is a challenge.  During normal collider operation, the fraction of single beam data is small and most decay data is sampled during few second pauses of top-up injections in physics runs. In other words, the data may not constrain all individual sensitivities for all accelerator conditions encountered on the training data. One countermeasure is to initialize the output neurons of the sensitivity networks well: by choosing the bias of the final dense layer in each sensitivity network $S_i$ such that the output of the layer equals the sensitivity measured during the last background study day. For $S_7$, the luminosity sensitivity model, we found that fixing the output to the sensitivity value extracted from the latest study day achieved most consistent results.   

\subsection{Hardware requirements}
Training the neural network on an AMD Ryzen 5 2600, a CPU with 6 cores at 3.40GHz clock speed, reaches an average of 16,400 samples trained per second. With a training dataset spanning three weeks, consisting of ~540,000 data points, this results in about $\SI{30}{\second}$ per epoch. A full training, consisting of 70 epochs, takes around $\SI{35}{\minute}$.
Using the trained model to make predictions on archived data on the same hardware results in 27,500 predicted samples per second.

\section{Delay optimization}

\begin{figure}[!ht]
\centering
\includegraphics[width=0.9\linewidth]{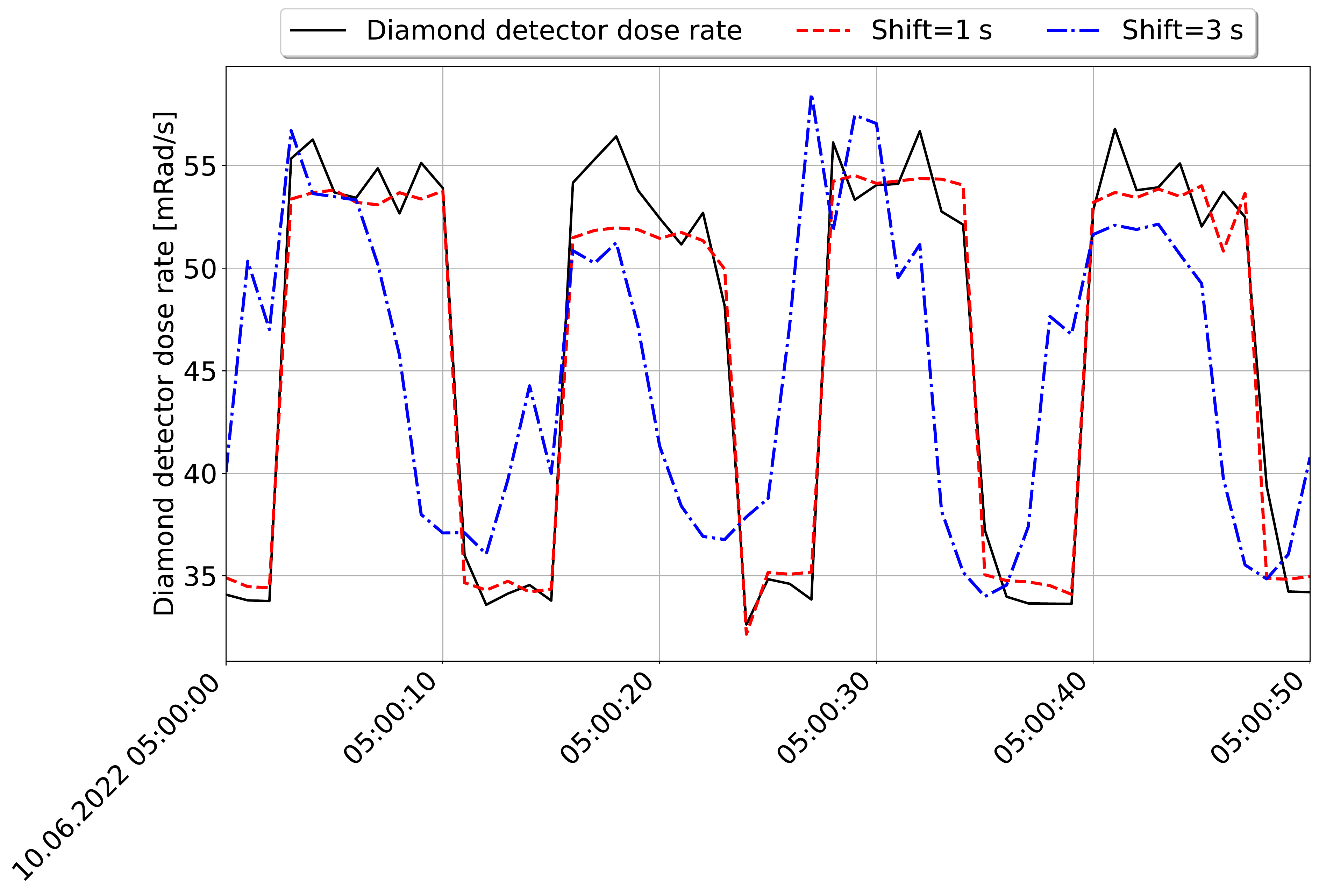}
\caption{\label{fig:different_shifts} Measured hit rate for a VXD diamond and the predictions of two BGNet models trained with  different delay shifts in data preprocessing. The shown time interval is from the 10.06.2022.}
\end{figure}

\begin{figure}[!ht]
\centering
\includegraphics[width=0.75\linewidth]{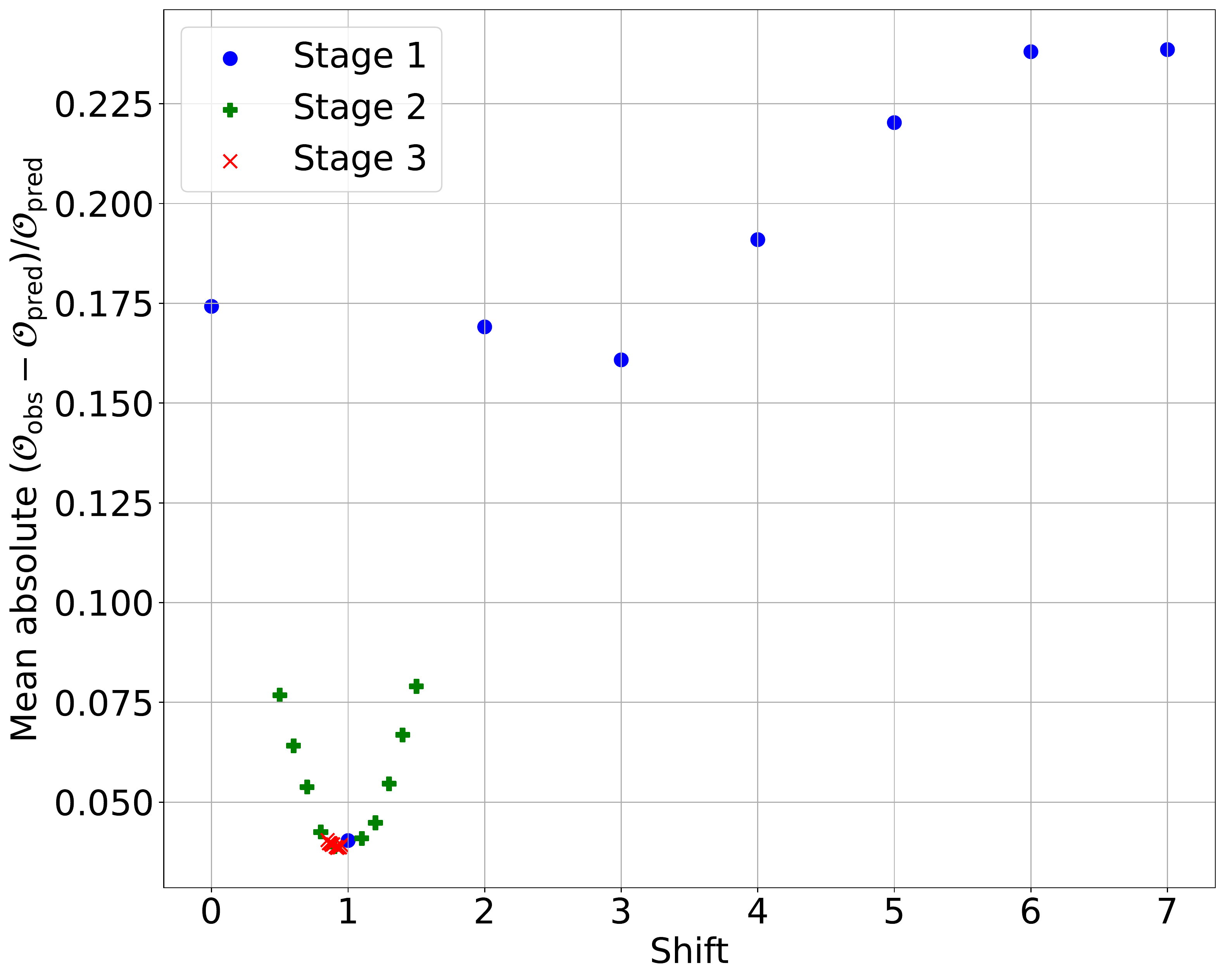}
\caption{\label{fig:delay_kernel_optimization}Network performance during physics runs dependent on the input delay shift of the VXD Diamond at $\phi=325$ in backwards direction mounted on the beam pipe}
\end{figure}

The delay correction is parameterized by a number called delay shift $d$. The fractional part of $d$ interpolates between shifting the input array $x$ by full seconds. For example, a delay shift of $2.4$ means we use the weighted average  $0.6\cdot x(t-2s) + 0.4\cdot x(t-3s)$ as input for estimating the backgrounds at time $t$.

The impact of the delay shift is especially noticeable during the top-up injections during physics runs. The pausing of top-up injections for multi second time intervals causes injection spikes in the measured hit rate. Figure \ref{fig:different_shifts} shows four such injection spikes recorded by the VXD beam abort diamond detector system~\cite{Bacher2021}. A delay shift between the measured hit rate and the input array, in particular the beam gate status variable, leads to a phase shift between the predicted and measured hit rate. This effect can be seen in Fig.~\ref{fig:different_shifts} where two BGNet models where trained for different values of the delay shift. 

The performance of a given delay shift can therefore be evaluated by the prediction error during physics runs. The performance metric is the mean absolute difference between predicted and measured hit rate, normalized to the measured hit rate. Using this characterization one can perform an optimization of the delay shift. This optimization is done in three stages. In the first stage, a grid search over a range of integer delay shifts is performed. In the second and third stage, the delay shift is further optimized by doing grid searches in smaller increments around the best result from the previous stage. The best performance is given by the delay shift with the smallest error on top-up injections. An illustration of such an optimization can be seen in Fig.~\ref{fig:delay_kernel_optimization}.

We find different optimal delay shifts for different Belle~II sub-detectors. The optimal delay shift is $\SI{1}{s}$ for the beam abort diamonds installed around the vertex detector (VXD), $\SI{5.5}{s}$ for the Central Drift Chamber (CDC) chamber current and $\SI{3.4}{s}$ for hit rates from Time-of-Propagation (TOP) detector photo multiplier tubes. We find no significant changes of the delay shifts during the run periods from 2020 to summer 2022.  

\section{Testing the background decomposition}

\begin{figure}[H]
    \centering
    \includegraphics[width=.7\linewidth]{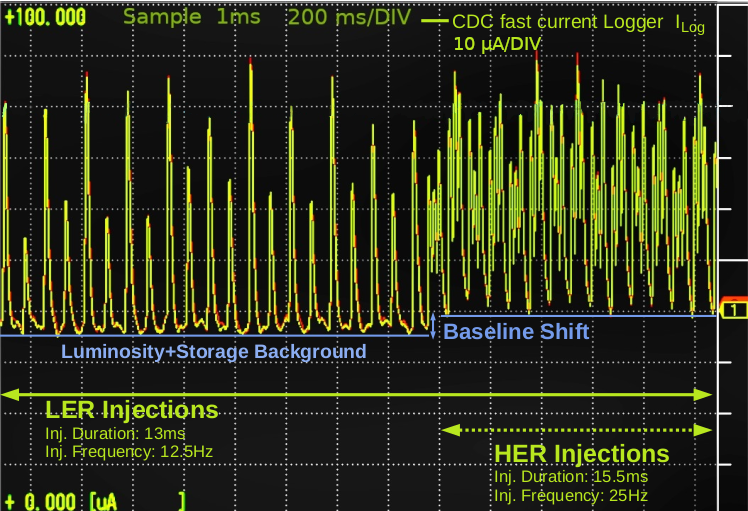}
    \caption{Fast CDC current logger resolves background spikes due to individual injections into the HER and LER. The figure shows multiple background spikes from injections into the LER followed by asynchronous injections into both the HER and LER during physics data taking. The timestamp of the data is 20:52:39 on June~12, 2022 (JST).}
    \label{fig:fast_current_monitor}
\end{figure}

The CDC~\cite{Abe:2010sj} in Belle~II provides a good example to experimentally study the quality of the background decomposition. 
The hit rate in the CDC volume is monitored online by the chamber current: the current over all wires per layer averaged over all 56 CDC layers. The chamber current is sampled with $\SI{1}{\hertz}$ frequency. In addition, the CDC has a fast current logger that samples the chamber current $I_{\text{Log}}$ of a sector of the fourth CDC layer at a kilohertz rate. This sampling rate is high enough to resolve spikes from individual top-up injections during physics runs. Figure.~\ref{fig:fast_current_monitor} shows an oscilloscope display of the fast current logger during a 3-second time window in June 2022. The figure shows multiple background spikes from injections into the LER at a repetition frequency of $\SI{12.5}{\hertz}$, injecting a charge of $\SI{1.16}{\nano\coulomb}$ per bunch. During this period, interleaved 1-bunch and 2-bunch injection into the LER was enabled and every second peak corresponds to the combined injection background from two bunches injected with a spacing of a few nanoseconds. This is followed by a period of injections into both the HER and LER during physics data taking. During that time, the HER injection follows a 1-bunch injection scheme at a repetition frequency of $\SI{25}{\hertz}$, injecting a charge of $\SI{3.21}{\nano\coulomb}$ per bunch. The achieved beam current during that time is $\SI{955}{\milli\ampere}$ ($\SI{1200}{\milli\ampere}$) for the HER (LER), and the peak luminosity is $\SI{3.46e34}{\centi\metre^{-2}.\second^{-1}}$.  After each injection, the chamber current decays during a few milliseconds to a baseline chamber current from combined storage and luminosity backgrounds. The baseline shift observed in Fig.~\ref{fig:fast_current_monitor} most likely results from not fully decayed injection spikes due to decay time and asynchronous injections into the LER and HER. The fast current logger provides the minimum chamber current $I^{\text{min}}_{\text{Log}}$ during one second to the EPICS Archiver Appliance. The minimum chamber current $I^{\text{min}}_{\text{Log}}$ serves as an estimate for the sum of luminosity and storage backgrounds. In order to compare it with the total CDC chamber current, it must be multiplied by a factor of four. This empirical factor arises from the fact that only a small sector of the CDC is taken into account for $I_{\text{Log}}$ measurements.

\begin{figure}[H]
    \centering
    \includegraphics[width=.6\linewidth]{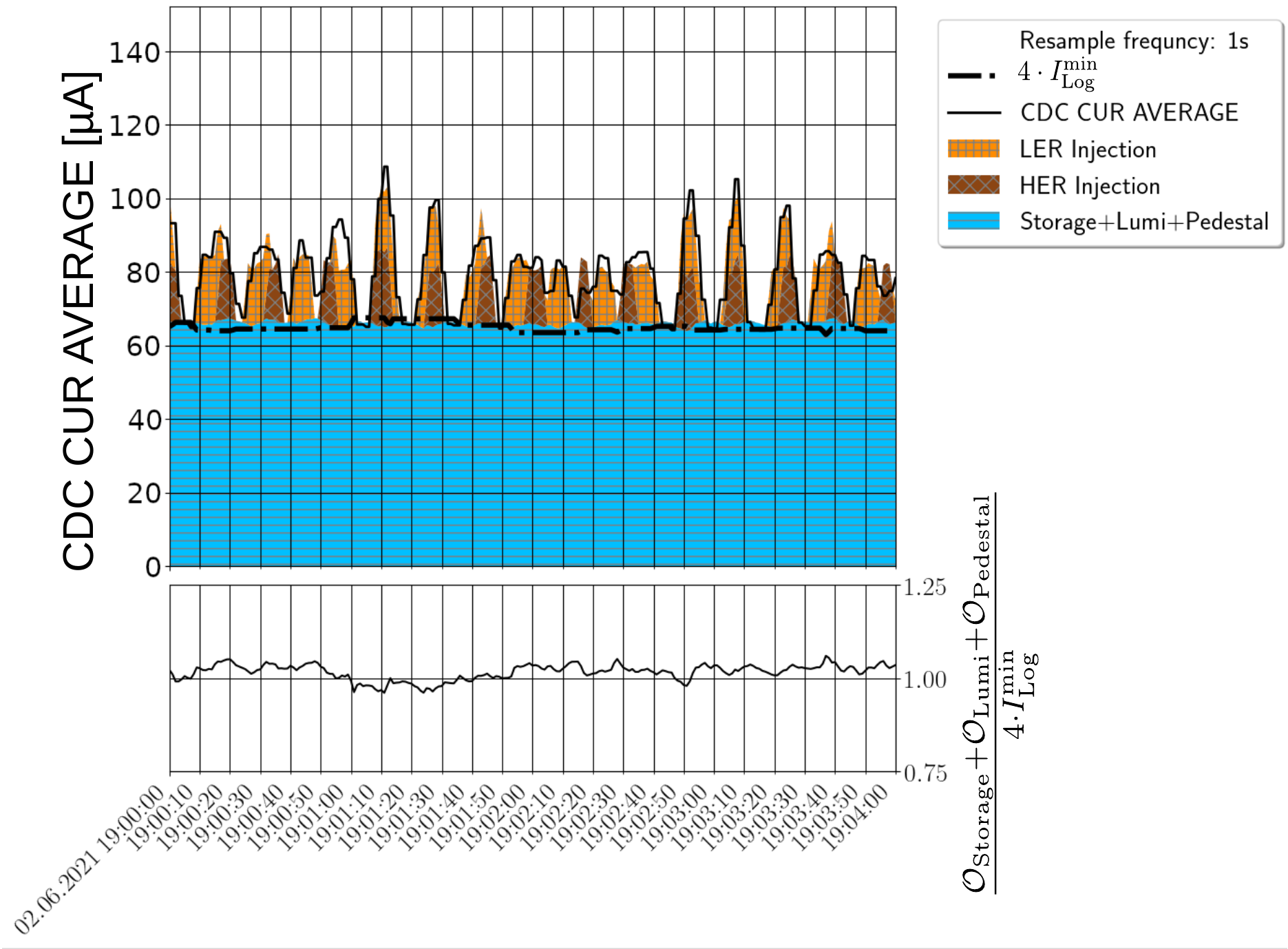}%
    \includegraphics[width=.4\linewidth]{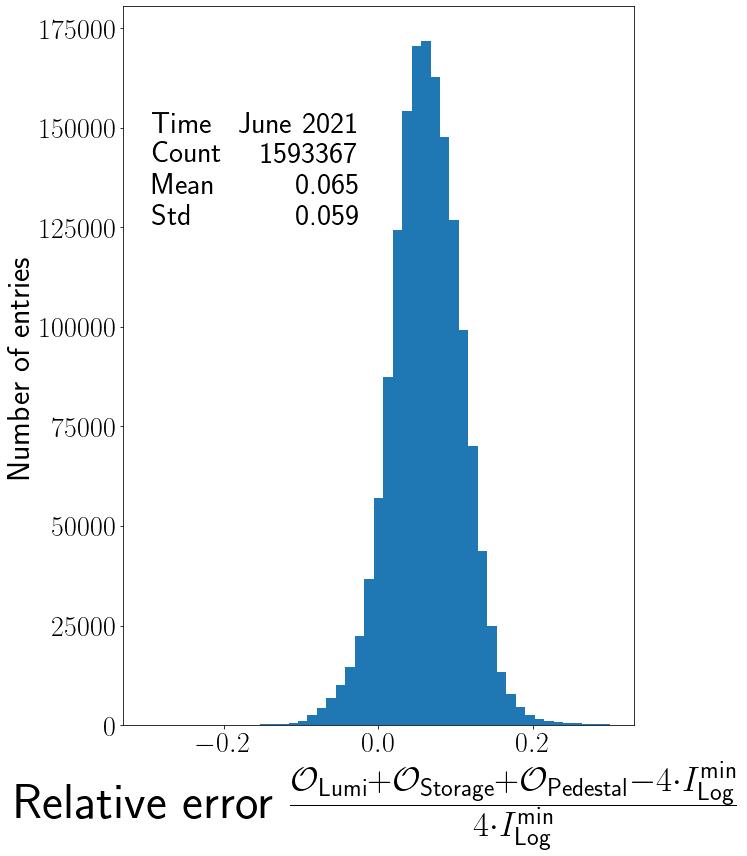}
    \caption{Left: Top-up injections in a 4~min time window. The storage, luminosity and pedestal components
    are combined and displayed together with the injection background components of HER and LER rings. The fast CDC current logger variable is used to gauge the ability of BGNet to differentiate between the sum of storage and luminosity backgrounds and injection backgrounds.
    Right: Histogram of the relative error between the fast CDC current logger variable and the predicted storage and luminosity backgrounds plus the pedestal over the month of June 2021.}
    \label{fig:top_up_logger}
\end{figure}

Figure~\ref{fig:top_up_logger} shows a comparison of $I^{\text{min}}_{\text{Log}}$ and the predictions of BGNet trained on the total CDC chamber current using training data covering June 2021. The figure shows the predicted injection backgrounds along with a sum over all storage and luminosity backgrounds for a $\SI{4}{\minute}$ time window during a physics run. The $I^{\text{min}}_{\text{Log}}$ agrees well with the BGNet prediction for the sum of storage and luminosity backgrounds. In other words, the BGNet model learned an accurate decomposition between injection and non-injection backgrounds even though it only used inputs sampled at $\SI{1}{\hertz}$ frequency. The histogram in Fig.~\ref{fig:top_up_logger} shows that the mean and the standard deviation of relative error during June 2021 are well below 10\%. The figure also confirms that injection backgrounds quickly decay to zero whenever injections are paused. The measured chamber current during such pauses provides an estimate for the sum of storage and luminosity backgrounds.

\section{Extrapolation with BGNet}
\begin{figure}
    \centering
       \includegraphics[width=1\linewidth]{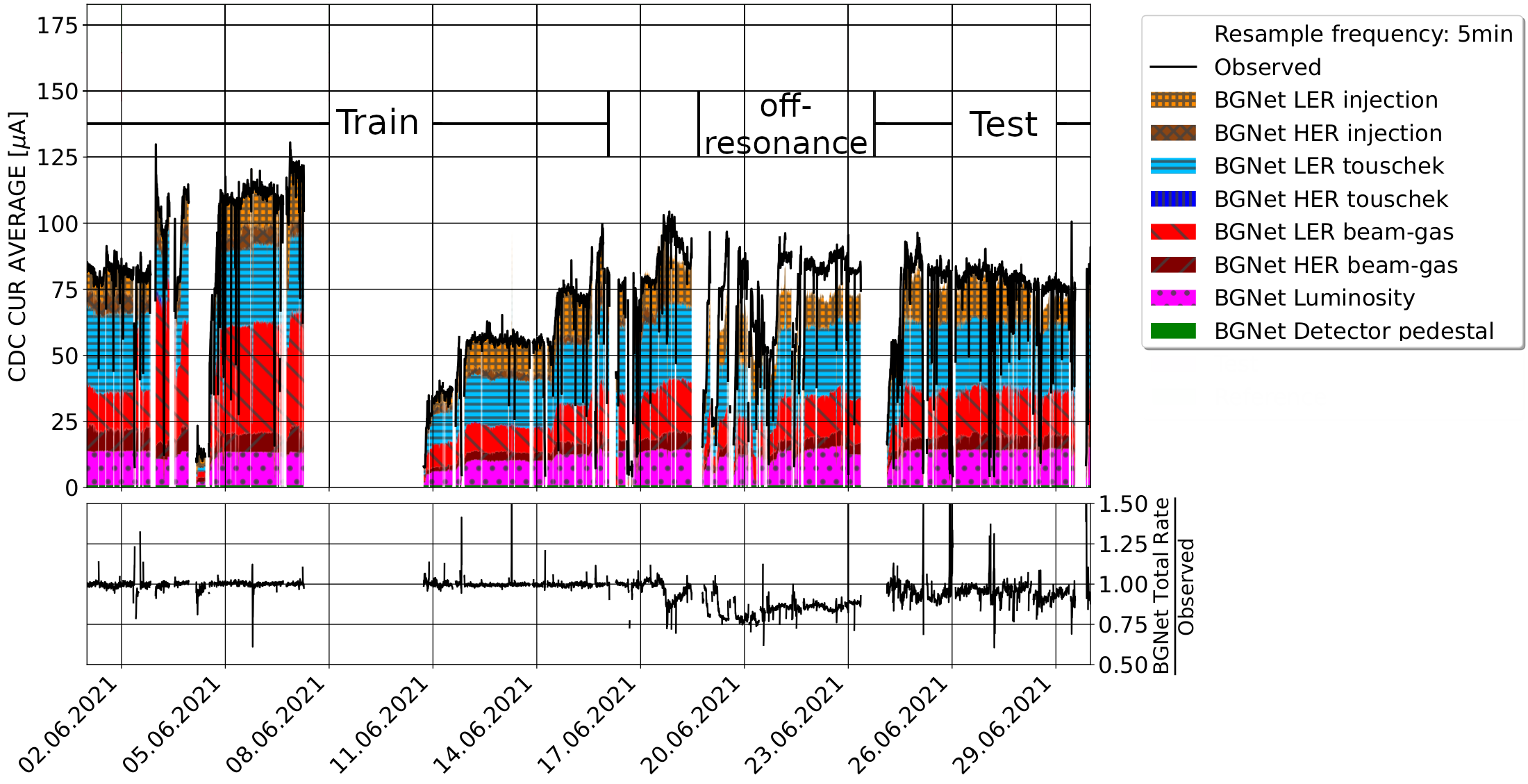}
    \caption{Stack plot showing the predicted decomposition and the measured CDC chamber current during June 2021. Training samples were drawn from June 1 to midnight of June 16 and the model was tested on the remainder of the month. Form June 19 to June 24 the collider conducted a Off-Resonance scan.}
    \label{fig:extrapolation_month}
\end{figure}

\begin{figure}[H]
\centering
    \begin{subfigure}[b]{0.85\textwidth}
       \includegraphics[width=1\linewidth]{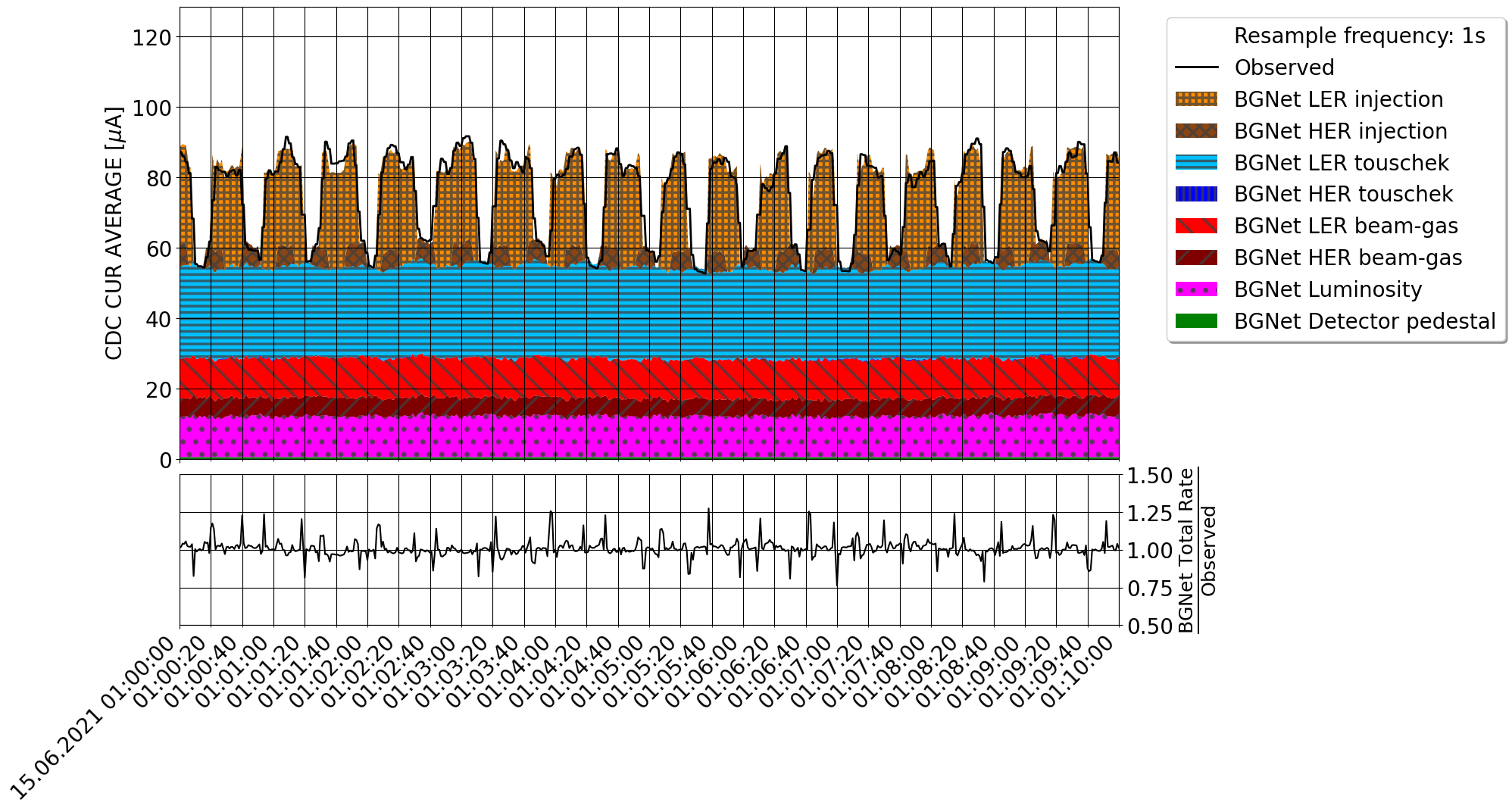}
    \end{subfigure}
    \begin{subfigure}[b]{0.85\textwidth}
       \includegraphics[width=1\linewidth]{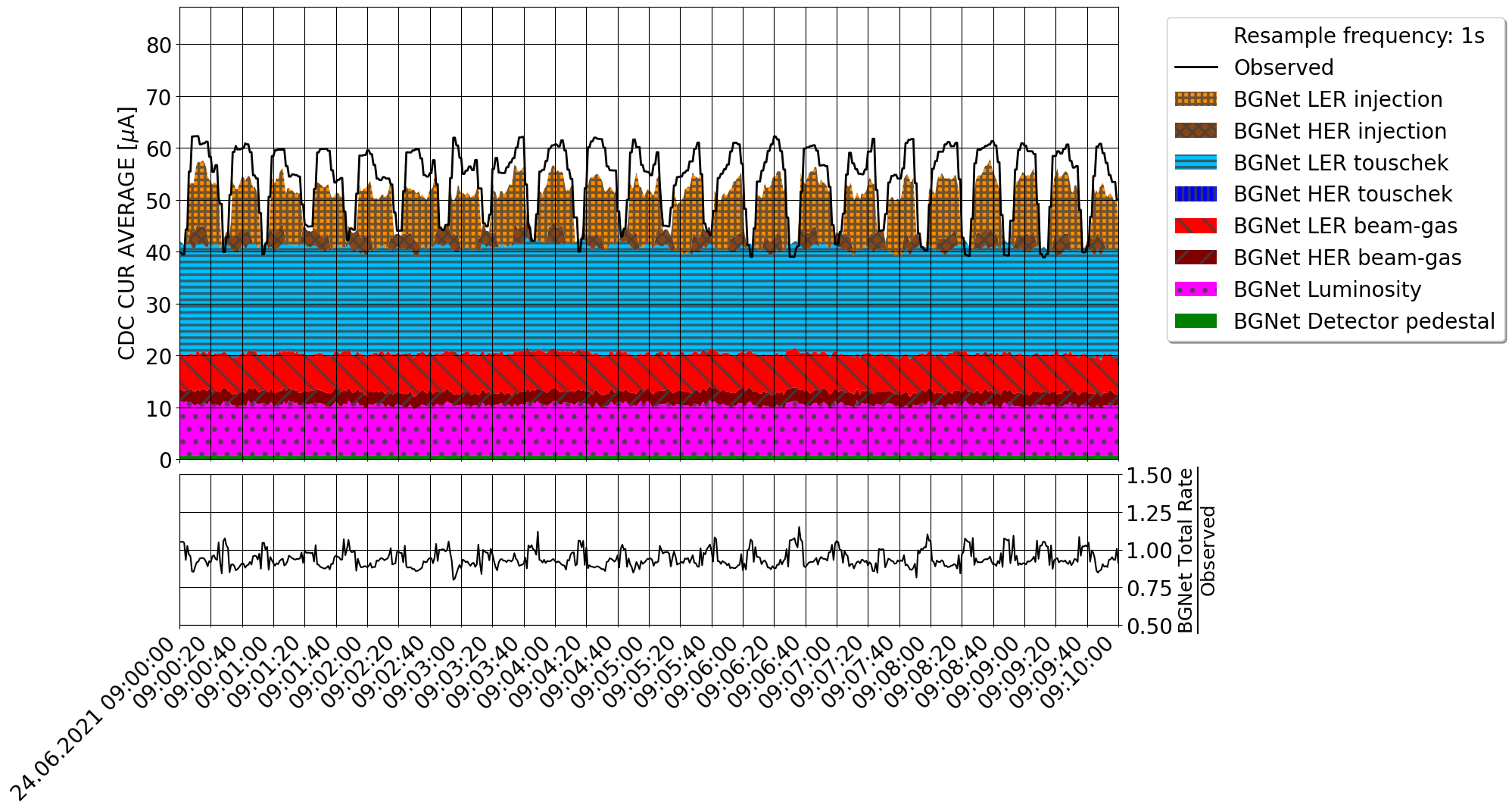}
    \end{subfigure}
    \caption{Comparison of measured and predicted background CDC hit rates during two time windows in June 2021. After training with archived data from the first half of June 2021, the model was applied to predict backgrounds on the remainder of June 2021. We show time windows in the training data and 8 days after the end of the training data.}
    \label{fig:extrapolation}
\end{figure}

In order to be useful as a real time diagnostic tool, BGNet needs to provide accurate predictions on incoming data beyond the time window used for training. Figure~\ref{fig:extrapolation_month} contains comparisons between BGNet predicted and measured hit rates for the CDC chamber current. BGNet was trained with archived hit rates on the first half of June 2021 at the nominal center-of-mass energy of $\SI{10.58}{GeV}$, including a background study conducted on June 16, 2021. The trained model was applied to predict the backgrounds during the full month containing an off-resonance energy scan from June 19 to 24 which was followed by data taking on the nominal center-of-mass energy. Table~\ref{tab:extrapolation_metrics} gives the mean and $\sigma_{68}$ of the relative prediction error separately for data from the first half of June, the Off-Resonance scan and the remainder of June. Background sensitivities are underestimated (biased) by 14\% during the Off-Resonance scan but the bias reduces to 4\% on test data once data taking at the nominal center-of-mass energy is resumed. Fig.~\ref{fig:extrapolation} shows two 10-minute time windows of physics  data taking on June 15 (inside the train time window) and June 24 (inside the test time window) with $\SI{1}{\hertz}$ frequency. The timing and duration of backgrounds from top-up injections is well described in both time windows. In between the injection spikes, the sum of storage backgrounds and luminosity background agrees well with the measured data. The main discrepancy is an underestimation of the background amplitude from top-up injections on June 24. A similar analysis for the off-resonance data shows that the observed bias of 14\% originates mostly from a underestimation of the HER injection background amplitude. There is an ongoing effort to improve predictions of injection backgrounds in the future by tuning the set of injection related input variables.

\begin{table}[H]
    \centering
    \begin{tabular}{c|c c}
         &  Mean $\frac{\mathcal{O}_{\text{obs}}-\mathcal{O}_{\text{pred}}}{\mathcal{O}_{\text{obs}}}$ & $\sigma_{68}$$\frac{\mathcal{O}_{\text{obs}}-\mathcal{O}_{\text{pred}}}{\mathcal{O}_{\text{obs}}}$\\[4pt]
         \hline
         Train& -0.0012 & 0.08\\
         Off-resonance& 0.14 & 0.13\\
         Test& 0.04 & 0.15
    \end{tabular}
    \caption{Performance metrics for BGNet on the three time windows (Train, Off-Resonance, Test) defined in Fig~\ref{fig:extrapolation_month}. The first column gives the mean relative prediciton error and the second column the range of the central 68\% of relative errors.}
    \label{tab:extrapolation_metrics}
\end{table}

\newpage
\section{Towards explaining backgrounds\label{sec:explaining_bgnet}}

BGNet learns surrogate models for the background sensitivities to all background types. So far, we provided case studies showing that the sensitivity models provide an accurate prediction of background levels. Here, we address the question if we can analyze the BGNet model to understand which input features are responsible for an observed change of the background level. To address this question, we look at archived data from the beginning of June 2022 where the origin of a background change is experimentally well known.  

\begin{figure}[!h]
\centering
\includegraphics[width=1.0\linewidth]{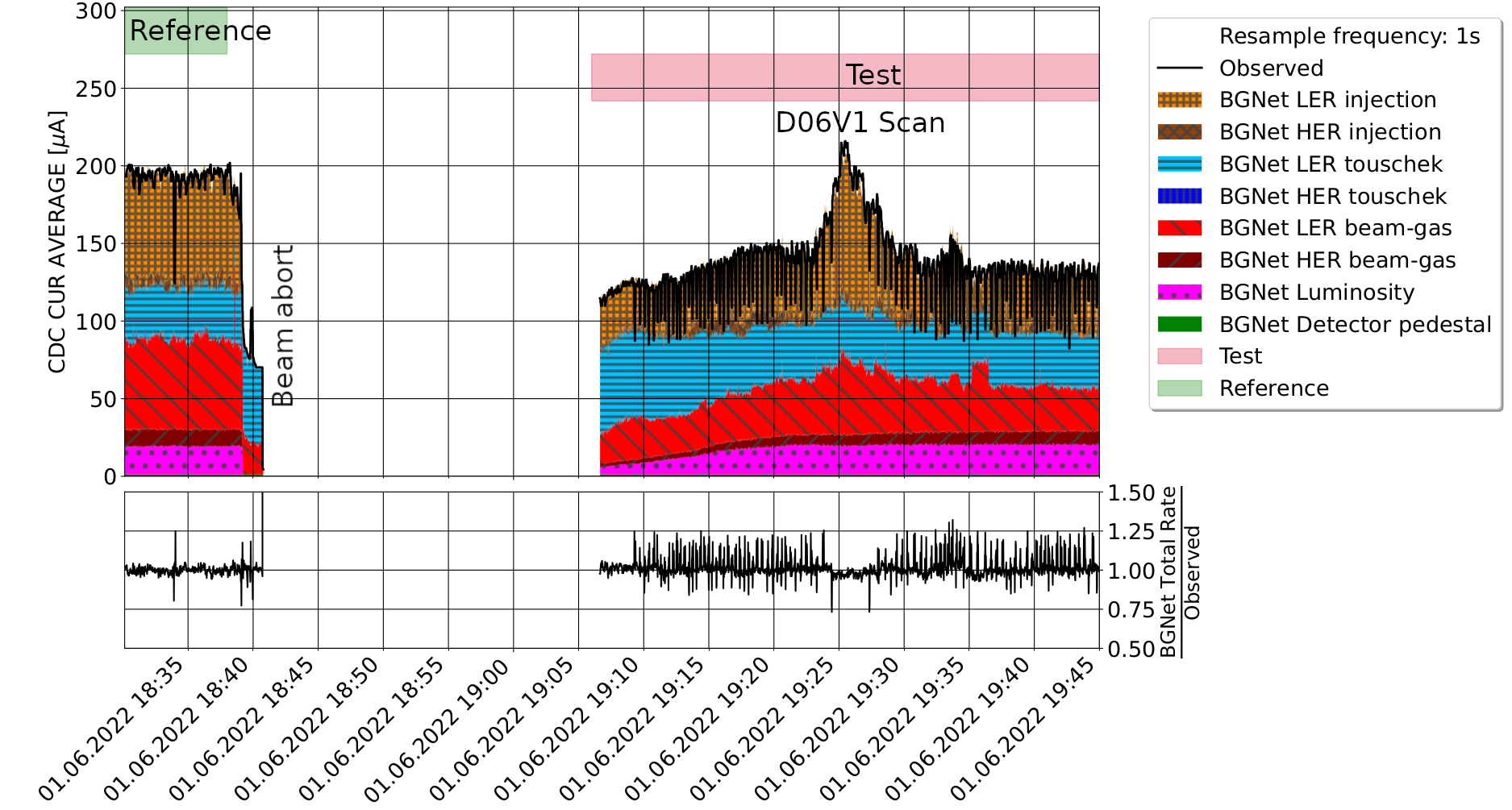}
\caption{\label{fig:stacked_plot_collimator_scan}Stack plot showing the predicted decomposition and the measured CDC chamber current. The peak of the chamber current corresponds to the opening of the collimator between 19:23 and 19:25, and closing between 19:25 and 19:30. } 
\end{figure}

Since BGNet already predicts the total hit rate as a sum over predictions for all background types, looking at the decomposition of background predictions already offers some insight. Figure~\ref{fig:stacked_plot_collimator_scan} shows  the background decomposition during the evening on June 1, 2022 made by a model trained on the data from May 27, 2022 to June 23, 2022.
A beam abort happened at 18:40.  During the next runs after the abort the measured CDC chamber current is much reduced. The predictions for the LER beam-gas and LER injection background show the same reduction. The figure also highlights a scan of the D06V1 collimator aperture used to experimentally locate the origin of the background reduction.   

Further information can be gained by applying feature attribution for each individual background sensitivity network $S_i(x)$.  As the sensitivity models are real valued and differentiable, we can directly apply the method of expected gradients~\cite{Erion}. The expected gradients attribution for the $j$th input feature of the sensitivity model $S_i$ is computed as

\begin{equation}
    \phi_{ij}(x) = \int_{x^{\star}} \left( (x_{j} - x_{j}^{\star}) \times \int_{\alpha=0}^{1} \frac{\partial S_i(x^{\star} + \alpha (x-x^{\star}))}{\partial x_j}d\alpha \right) p(x^{\star})dx^{\star},
    \label{eq:path_explain_formula}
\end{equation}
where $x$ is the test sample to be explained and $x^\star$ is a reference sample drawn from any user-defined distribution $p(x^{\star})$ over the data. In order to explain changes background hit rates over time, we are using a uniform distribution over a reference time window. The inner integral in Eq.~\ref{eq:path_explain_formula} integrates the partial derivative of the sensitivity along a straight line connecting the point $x$ with the reference point $x^{\star}$ in the feature space. The attribution value quantifies how much the selected feature contributes to the difference between $S_i(x)$ and the expectation value $<S_i>$ over the distribution $p(x^{\star})$ of reference samples. We use the python library  \textit{Path Explain}~\cite{janizek2020explaining} to compute attribution values.

\begin{figure}[!h]
\centering
\includegraphics[width=1\linewidth]{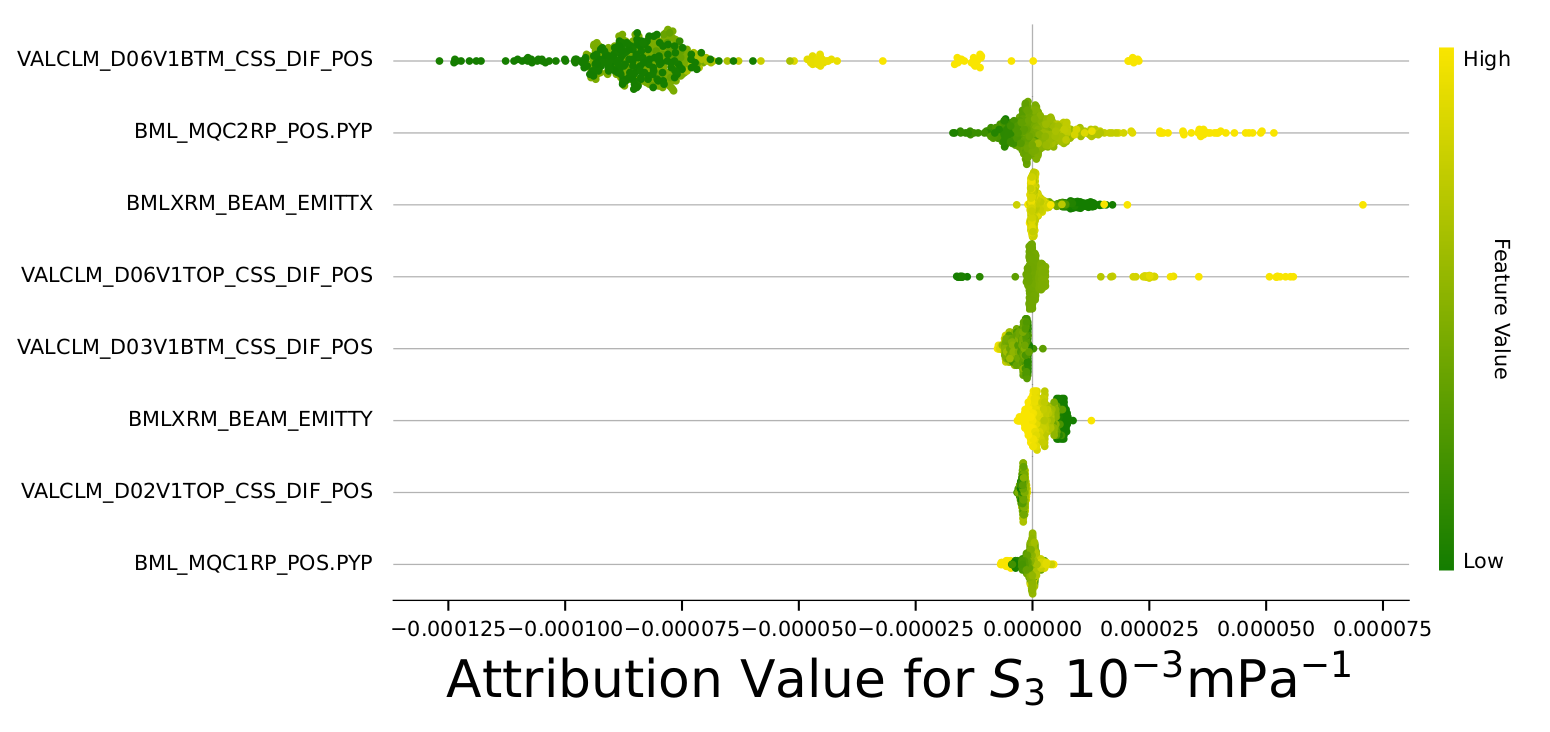}
\caption{\label{fig:attribution_plot_collimator_scan} Summary of the top eight features and their attribution values for the LER beam-gas sensitivity. Input variables 'VALCLM\_*\{TOP,BTM\}\_CSS\_DIF\_POS' refer to the monitored aperture of the top or bottom jaw of vertical collimators in different sections of the LER ring. Variables 'BMLXRM\_*\_EMITT\{X,Y\}' refer to readings from the X-ray monitor for the LER beam emittance. Variables 'BML\_*\_POS.PYP' refer to vertical beam position in different parts of the superconducting final focusing magnets.}
\end{figure}

This method gives a handle to understand why the model predicts a reduction in the LER beam-gas hit rate in the runs after the beam abort.  We use a uniform distribution $p(x^{\star})$ over a time window before the beam abort. We compute attributions using test data points sampled uniformly from runs after the beam abort.  Figure~\ref{fig:attribution_plot_collimator_scan} shows a summary of the attributions computed for the LER beam-gas sensitivity  $S_3(x)$ model  using Eq.~\eqref{eq:path_explain_formula}. Each point represents a test sample, its color gives the value of the feature and its position on the x-axis gives the attribution value. The attribution value tells us how much this feature contributed to the reduction of the LER beam-gas sensitivity after the beam abort compared to the sensitivity before the beam abort. The features are ranked according to the mean absolute attribution value over all tested data points.

\begin{figure}[!h]
\centering
\includegraphics[width=0.7\linewidth]{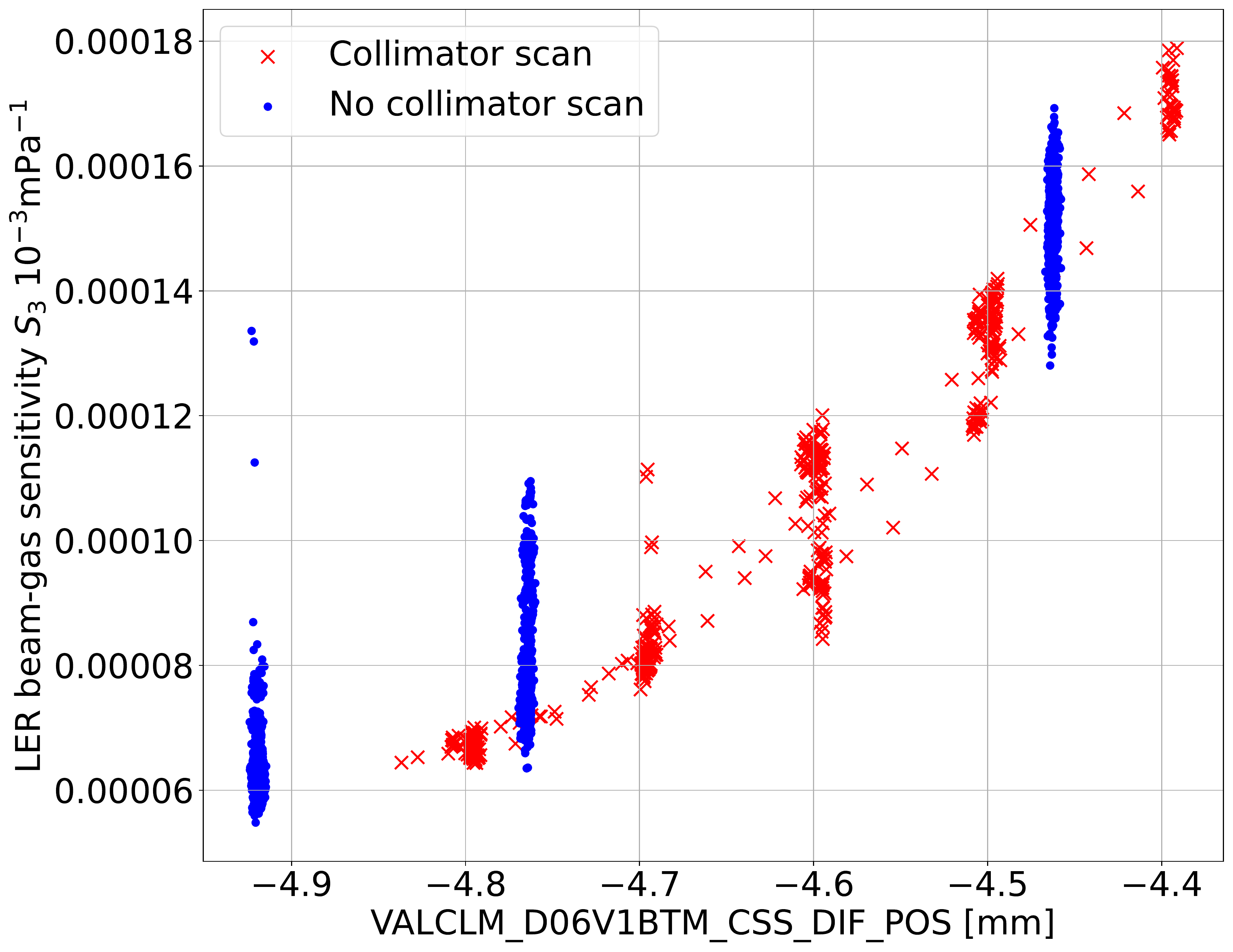}
\caption{\label{fig:scatter_plot_collimator_scan}Scatter plot of the predicted LER beam-gas sensitivity and the aperture of the bottom jaw of the D06V1 collimator on samples from the 01.06.22 between 18:20 and 19:45.}
\end{figure}

Figure~\ref{fig:attribution_plot_collimator_scan} shows that the LER beam-gas sensitivity model largely attributes the reduction in the sensitivity to a change of the aperture of the collimator bottom jaw: the LER beam-gas background was reduced by opening of the D06V1 collimator. The D06V1 collimator scan  further reinforces this conjecture, as adjustments in the aperture coincide with large changes in the measured hit rate. Figure~\ref{fig:scatter_plot_collimator_scan} shows the behaviour of the predicted beam-gas sensitivity for different collimator apertures.

\section{Conclusion}

We introduced a novel neural network based model, BGNet, to predict the beam-induced background hit rate in Belle~II sub-detectors. 
The neural network model architecture uses the particle scattering formulae and domain knowledge to explicitly decompose the hit rate of the Belle~II detector into a sum of contributions for different background types.
The BGNet model predicts the  sensitivity  to a background type by a fully connected neural network using a selection of diagnostic variables provided by  the collider subsystems as input. All input variables are archived on the EPICS Archiver Appliance at a $\SI{1}{\hertz}$ sample frequency. The role of these neural networks is to provide a flexible parametrization of how different SuperKEKB collider parameters affect the Belle~II sensitivity to a specific type of background.
After training the model on archived samples from the last few weeks of collider operation, the model is able to detect the most crucial parameters to be adjusted for background mitigation and collider performance improvement.

A strength of the BGNet model is its ability to disentangle the contribution to the hit rate from storage beam losses against contributions from  top-up injections into the HER or LER. This was achieved  by modulating the predicted injection background amplitude by the beam gate status input variable that indicates if top-up injections are paused. In the case of the CDC, we could explicitly test our decomposition based on $\SI{1}{\hertz}$ inputs with a direct measurement of the fast CDC current logger, a detector measuring the background induced chamber current in a sector of the CDC at a sample rate of 1ms.

We provide a case study to show that BGNet is able to learn which moveable collimators affect the sensitivity to storage and top-up injection induced beam losses. In this example, we demonstrated how feature attribution methods can be used to find the most relevant change of an input feature for explaining a change on the model prediction. The example illustrates how machine learning can provide  insights for background mitigation and control. We plan to integrate the neural network based background decomposition into the SuperKEKB background monitor panel. A display showing the real-time background decomposition at Belle~II can be used by SuperKEKB machine-operators to move collimators and tune top-up injections into SuperKEKB.

\section*{Acknowledgements}

The authors are grateful to Belle~II and SuperKEKB colleagues for their hard work and contribution. We acknowledge the financial support by the Federal Ministry of Education and Research of Germany. This work was supported by the U.S. Department of Energy (DOE) via Award Number DE-SC0010504 and via U.S. Belle~II Operations administered by Brookhaven National Laboratory (DE-SC0012704).
\bibliography{mybibfile}

\end{document}